\documentstyle[aps,prd,times,psfig]{revtex}

\begin{document}
\draft

\title{Properties of $r$ modes in rotating magnetic
	neutron stars.\\
	II. Evolution of the $r$ modes and stellar
	magnetic field.}

\author{Luciano~Rezzolla}
\address{SISSA, International School for Advanced Studies,
         Trieste, Italy.}
\address{INFN, Department of Physics, University of Trieste, Italy.}

\author{Frederick L. Lamb$^{\rm a,b,c}$,  
    	Dragoljub Markovi\'c$^{\rm a,b}$, and 
	Stuart L. Shapiro$^{\rm a,b,c,d}$}

\address{$^{\rm a}$Center for Theoretical Astrophysics,
         University of Illinois at Urbana-Champaign,
         Urbana, Illinois.}
\address{$^{\rm b}$Department of Physics, University of
         Illinois at Urbana-Champaign, Urbana, Illinois.}
\address{$^{\rm c}$Department of Astronomy, University of
         Illinois at Urbana-Champaign, Urbana, Illinois.}
\address{$^{\rm d}$NCSA, University of Illinois at
         Urbana-Champaign, Urbana, Illinois.}

\medskip

\date{\today}
\maketitle

\begin{abstract}
\noindent The evolution of the \hbox{$r$-mode}
instability is likely to be accompanied by secular
kinematic effects which will produce differential
rotation with large scale drifts of fluid elements,
mostly in the azimuthal direction. As first discussed
in~\cite{rls00}, the interaction of these secular
velocity fields with a pre-existing neutron star magnetic
field could result in the generation of intense and large
scale toroidal fields. Following their derivation in the
companion paper~\cite{rlms01a}, we here discuss the
numerical solution of the evolution equations for the
magnetic field. The values of the magnetic fields
obtained in this way are used to estimate the conditions
under which the \hbox{$r$-mode} instability might be
prevented or suppressed. We also assess the impact of the
generation of large magnetic fields on the gravitational
wave detectability of $r$-mode unstable neutron
stars. Our results indicate that the signal to noise
ratio in the detection of gravitational waves from the
\hbox{$r$-mode} instability might be considerably
decreased if the latter develops in neutron stars with
initial magnetic fields larger than $10^{10}$ G.
\end{abstract}


\section{Introduction}
\label{intro}

	Hot, newly born and fast spinning neutron stars
are among the best candidates for triggering an $r$-mode
instability. These candidates, however, are also likely
to be threaded with intense magnetic fields in the range
$10^{11}-10^{13}$~G, that are usually neglected when
discussing the onset and growth of the $r$-mode
instability. As in our introductory paper~\cite{rls00},
we here focus our attention on the presence of these
magnetic fields. In particular, we develop in more detail
the evolutionary scenario outlined in~\cite{rlms01a}
(hereafter paper~I), where we have argued that the
kinematical properties of the $r$-mode oscillations will
produce secular velocity fields that couple to any
pre-existing magnetic field and result in a net
amplification of the latter.

	Using the set of induction equations derived in
paper~I, we here present the results of their numerical
integration and show that the exponential growth of the
mode amplitude is also accompanied by an exponential
growth of a toroidal magnetic field. By the time that the
instability has reached saturation, the newly generated
toroidal field might become comparable with or larger
than the seed poloidal magnetic field if the saturation
amplitude is of order unity. The generation of magnetic
field is accompanied by a conversion of the energy of the
mode into magnetic energy. By computing the growth rate
of the magnetic field, we can calculate the rate of mode
energy loss to magnetic energy and estimate the strength
of the magnetic field which would either prevent the
onset of the instability or suppress its evolution. In
our calculations we consider separately the case in which
the stellar core is composed of a normal neutron fluid
from the case in which the core is a Type~II
superconductor.

	The generation of magnetic field also decreases
the amount of energy which can be radiated to infinity in
the form of gravitational waves and we here discuss how
the expectations for detectability of $r$-mode unstable
neutron stars through gravitational radiation should be
modified in terms of the initial magnetic field in the
star and of the mode amplitude at saturation.

	The following is a brief outline of the contents
of the paper. In Section~\ref{rmi_evol} we present our
model for the evolution of the \hbox{$r$-mode}
instability in a magnetic neutron star. In particular, we
first discuss the background evolution of the star's
angular velocity and of the mode amplitude, and then
present our calculations for the evolution of the
magnetic fields. Next, we consider in Section~\ref{iomf}
the impact of the newly generated magnetic fields on the
evolution of the instability, estimating the critical
magnetic fields for the prevention of the instability or
its suppression.  Section~\ref{gwd} is devoted to a
revision of the standard picture for the gravitational
wave detectability of the \hbox{$r$-mode} instability for
a magnetic neutron star. Finally, conclusions and
perspectives for future developments are presented in
Section~\ref{conclusions}.

\section{Evolution of the $r$-mode instability in a
	magnetic neutron star}
\label{rmi_evol}

	The description of the fully developed character
of the instability is a non trivial task and, at present,
several important aspects of the instability still need
clarification. A self-consistent calculation of the
problem requires fully nonlinear magnetohydrodynamics as
well as General Relativity. This is beyond the scope of
this paper. However, some important conclusions can be
drawn by considering the $r$-mode instability and the
generation of magnetic fields as distinct processes,
evolving independently. This is equivalent to assuming
that the {\sl nature} of the $r$-mode oscillations is
unaffected by the generation of large magnetic fields and
can, at any time, be described by the perturbative
expressions [cf. Eq. (1) in paper~I]. Within this
framework, we can then specify a model for the evolution
of the \hbox{$r$-mode} instability and couple it with the
evolution of the magnetic field generated by the secular
effects produced in \hbox{$r$-mode} oscillations, and
discussed in detail in paper~I.

\subsection{Numerical model of $r$-mode evolution}
\label{nmrw}

	In order to provide a direct comparison with
results presented in the literature we have made use of
the phenomenological model for the evolution of the
instability presented by Owen et al.~\cite{oetal98} in
the case of a rotating neutron star with zero magnetic
field. In this model, the evolution of the
\hbox{$r$-mode} instability is assumed to have three
phases. The initial phase is the one during which any
infinitesimal axial perturbation will grow exponentially
over the timescale $\tau_{_{\rm GW}}$, set by the
gravitational radiation-reaction. For an $\ell = m = 2$
mode\footnote{Hereafter we will only consider modes with
$\ell=m$.} and a neutron star initially rotating at the
``break-up'' limit $\Omega_{_{\rm B}}$, at a temperature
$\sim 10^{10}$ K (and for a number of different equations
of state), this timescale has been estimated to be of the
order of a few tens of seconds. The exponential growth is
followed by an intermediate phase during which the
amplitude of the mode is expected to reach its saturation
value $\alpha_{\rm sat}$, and the star is progressively
spun down as a result of the angular momentum loss via
gravitational waves. This phase has been estimated to be
of the order of about one year if conventional cooling
rates and viscosity estimates are used. The final phase
of the evolution is the one in which viscous dissipative
effects dominate the radiation-reaction forces and the
$r$ modes start to be damped out.

	Phenomenological expressions for the time
evolution of the mode amplitude $\alpha$ and of the
star's angular velocity $\Omega$ can be derived by
requiring that the loss rates of the energy in the mode
and of the star's total angular momentum are the same as
those recorded at infinity in the form of gravitational
waves. The ordinary differential equations governing this
evolution have been first derived in~\cite{oetal98} and
can be synthesized as\footnote{Note that our expression
for $d\Omega/dt$ after saturation differs from the one
given in~\cite{oetal98}, where it was assumed that
$\alpha_{\rm sat}$ is a constant and it is always
${\mathcal O}(1)$, independently of the actual value of
$\alpha_{\rm sat}$.}

\vbox{
\begin{eqnarray}
&{\rm Before\ Saturation \hskip 5.0truecm} 
&{\rm After\ Saturation,\ Before\ Decay }
\nonumber \\ 
\label{adot_bs}
&\hskip -4.0truecm
\displaystyle{ {d\alpha\over dt}=-{\alpha\over\tau_{_{\rm GW}}}
	-{\alpha\over\tau_{_{\rm V}}}
	\left({1-\alpha^2Q\over 1+\alpha^2Q}\right)\ , }
&\hskip 1.0truecm
{d\alpha\over dt}=0\ ; \hskip 1.0truecm \alpha = \alpha_{\rm sat} \ ,
\\ \nonumber \\
\label{odot_bs}
&\hskip -5.0truecm
\displaystyle{ {d\Omega\over dt}=-{2\Omega\over \tau_{_{\rm V}}}
	\left({\alpha^2 Q\over 1+\alpha^2 Q} \right)\ , }
&\hskip 1.0truecm
{d\Omega\over dt} = {2\Omega\over \tau_{_{\rm GW}}}
	\left(
	\frac{Q\,\alpha^2_{\rm sat}}{1-Q\,\alpha^2_{\rm sat}} 
	\right) \ ,
\end{eqnarray}}
\noindent where ${\tau}_{_{\rm V}}$ is the viscous
timescales comprising both bulk and shear viscosity
(see~\cite{oetal98} for a definition) and $Q$ is a
constant [see Eq. (25) of paper~I for a definition].

\begin{figure}[htb]
\begin{center}
\leavevmode
\psfig{file=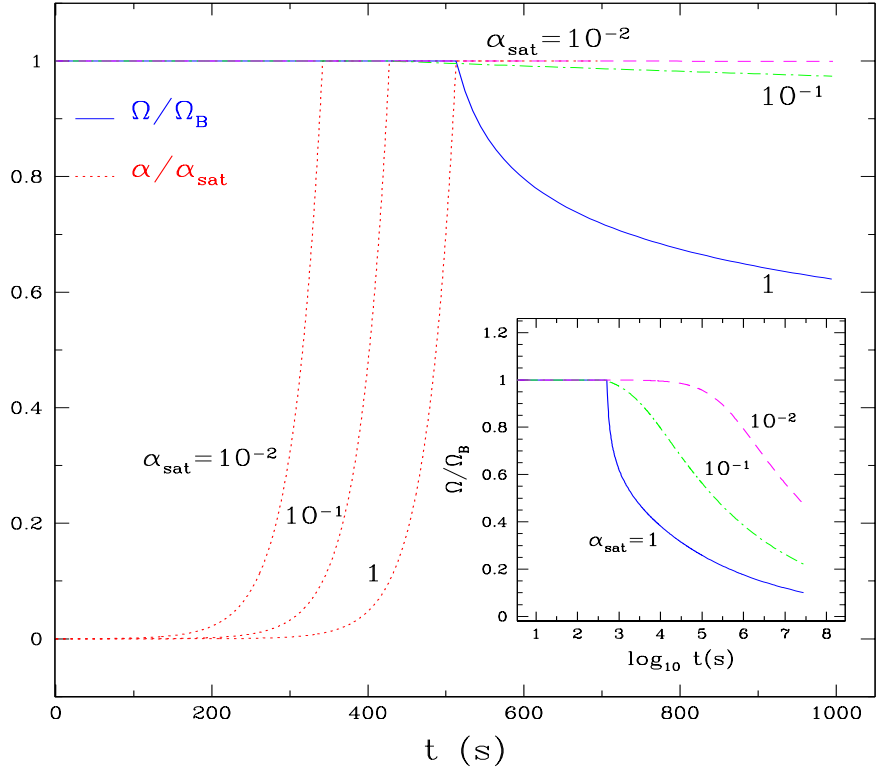,angle=0,width=13.0truecm,height=11.0truecm}
\end{center}
\caption[]{\label{fig1} Time evolution of the mode
amplitude (dotted lines) and of the star's angular
velocity (solid lines) normalized to the saturation value
$\alpha_{\rm sat}$ and the break-up limit $\Omega_{_{\rm
B}}$, respectively. Different curves refer to different
values of the saturation amplitude. The small inset shows
the decay of the star's angular velocity on a logarithmic
time scale.  }
\end{figure}

	The numerical solutions to equations
(\ref{adot_bs}) and (\ref{odot_bs}) are shown in
Fig.~\ref{fig1}, where we have plotted the time evolution
of the mode amplitude (dotted lines) and of the star's
angular velocity (solid lines), normalized to the
saturation value and the break-up limit respectively. The
main panel of Fig.~\ref{fig1} shows the initial phases of
the amplitude growth and angular velocity decay and is,
for this reason, shown on a linear temporal
scale. Different solid curves refer to different values
of the saturation amplitude. Note that the decrease in
the star's angular velocity and that its value after one
year depends sensitively on the value used for
$\alpha_{\rm sat}$ and becomes very small only for
$\alpha_{\rm sat}=1$. This is apparent from the inset
which shows the decay of the star's angular velocity on a
(longer) logarithmic time scale.

\begin{figure}[htb]
\begin{center}
\leavevmode
\psfig{file=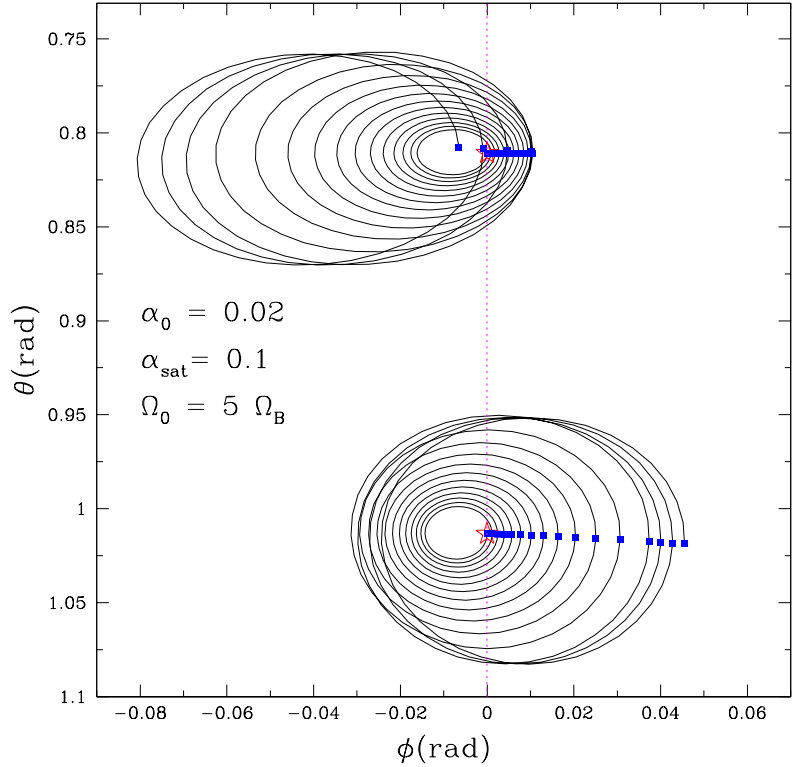,angle=0,width=11.0truecm}
\caption[]{\label{fig2} Projected trajectories
$\theta(t)\sin\theta(t)\cos\phi(t)$, and
$\phi(t)\sin\theta(t)\cos\phi(t)$ of two fiducial fluid
elements as seen in the corotating frame. While the fluid
elements are allowed to oscillate, the amplitude of the
mode increases from an initial value $\alpha_0 = 0.02$ to
a saturated one $\alpha_{\rm sat}=0.1$. The initial
positions are indicated with stars, while the positions
at the end of each period are indicated with filled
squares. Note that we have set $\Omega=5\Omega_{_{\rm
B}}$; this is to have a growth time which is larger but
comparable with the oscillation period and to show
schematically the effects of a increasing mode amplitude
on the secular motions.}
\end{center}
\end{figure}

	Next, we consider the effect of an increasing
mode amplitude on the kinematics of \hbox{$r$-mode}
oscillations. For doing this we compute the equations of
motion allowing the mode amplitude to vary as in Eq.
(\ref{adot_bs}) in the case before saturation.
Fig.~\ref{fig2} shows the effect of a variable mode
amplitude on the trajectories of two fiducial fluid
elements following $r$-mode oscillations with $\ell = 2$.
We concentrate on two fluid elements which have the same
initial longitude $\phi_0=0$ and latitude $\sim 50$
degrees, with the mode having an initial amplitude
$\alpha_0 = 0.02$, which then saturates at $\alpha_{\rm
sat} = 0.1$. The initial positions are indicated with
stars, while the positions at the end of each oscillation
are indicated with filled squares. Note that in order to
have a growth time which is larger but comparable with
the oscillation period, we have artificially set
$\Omega=5\Omega_{_{\rm B}}$ and the trajectories shown in
Fig.~\ref{fig2} should therefore be considered only as
indicative. 
	
	When compared with the corresponding trajectories
shown in Fig.~\ref{fig1} of paper~I, they show that the
effect of an exponentially increasing mode amplitude is
that of exponentially {\it increasing} the azimuthal and
polar drifts between oscillations. As discussed in
paper~I when deriving the expression for the azimuthal
drift [cf. Eq. (6) of paper~I], one should expect secular
effects to emerge every time that the velocity field at
the position of the fluid element varies as it
moves. This can happen either because the fluid velocity
has a dependence on the position of the fluid element (as
in the case of constant amplitude oscillations) or
because the amplitude of the fluid velocity is changing
(as during the growth of the instability). As a result, a
mode amplitude which is exponentially growing in time
will produce a secular motion, independently of whether
the modes have reached nonlinear amplitudes.

	In the following Section, where we discuss the
numerical evolution of the magnetic field as a result of
the secular velocity field, it will be possible to
distinguish the contribution to the field generation
coming from the exponential growth phase of the mode from
the one resulting from the secular drift at mode
saturation.

\subsection{Evolution of the stellar magnetic field}
\label{nr}

	We here present results from the solution of the
induction equations as obtained with both the Lagrangian
method [cf. Eq. (14) of paper~I] and with the
orbit-average method [cf. Eq. (16) of paper~I] discussed
in paper~I. The computations have been performed for a
number of different values of the saturation amplitudes,
for modes with $\ell=2, 3, 4$, and with an initial
dipolar magnetic field [cf. Eq. (8) of paper~I].

	From a numerical point of view, the Lagrangian
and orbit-average approaches have a common strategy. Both
of them, in fact, solve the equations of motion
[cf. eqs. (3) of paper~I] for a number of fiducial fluid
elements suitably distributed on an isobaric surface of
the star and calculated for several different mode
numbers. In the Lagrangian approach, in particular, the
new positions after each oscillation are used to
reconstruct the strain tensor $S_{jk}=\partial {\tilde
x}^{j}(t)/\partial {x}^{k}(t_0)$ from which the new
magnetic field components are computed. Conversely, in
the numerically more expensive Eulerian approach, the
secular polar and azimuthal velocities ${\tilde v}^{\hat
{\theta}}$, ${\tilde v}^{\hat {\phi}}$ are computed after
each oscillation period and their values transformed onto
a finite difference grid corotating with the star, over
which the induction equations are then solved. In this
latter case, a second-order Lax-Wendroff evolution scheme
is used to minimize artificial dissipative effects and
accurately estimate the field growth rate.

\begin{figure}[htb]
\begin{center}
\leavevmode
\hbox{
\hspace{0.125truecm}
\psfig{file=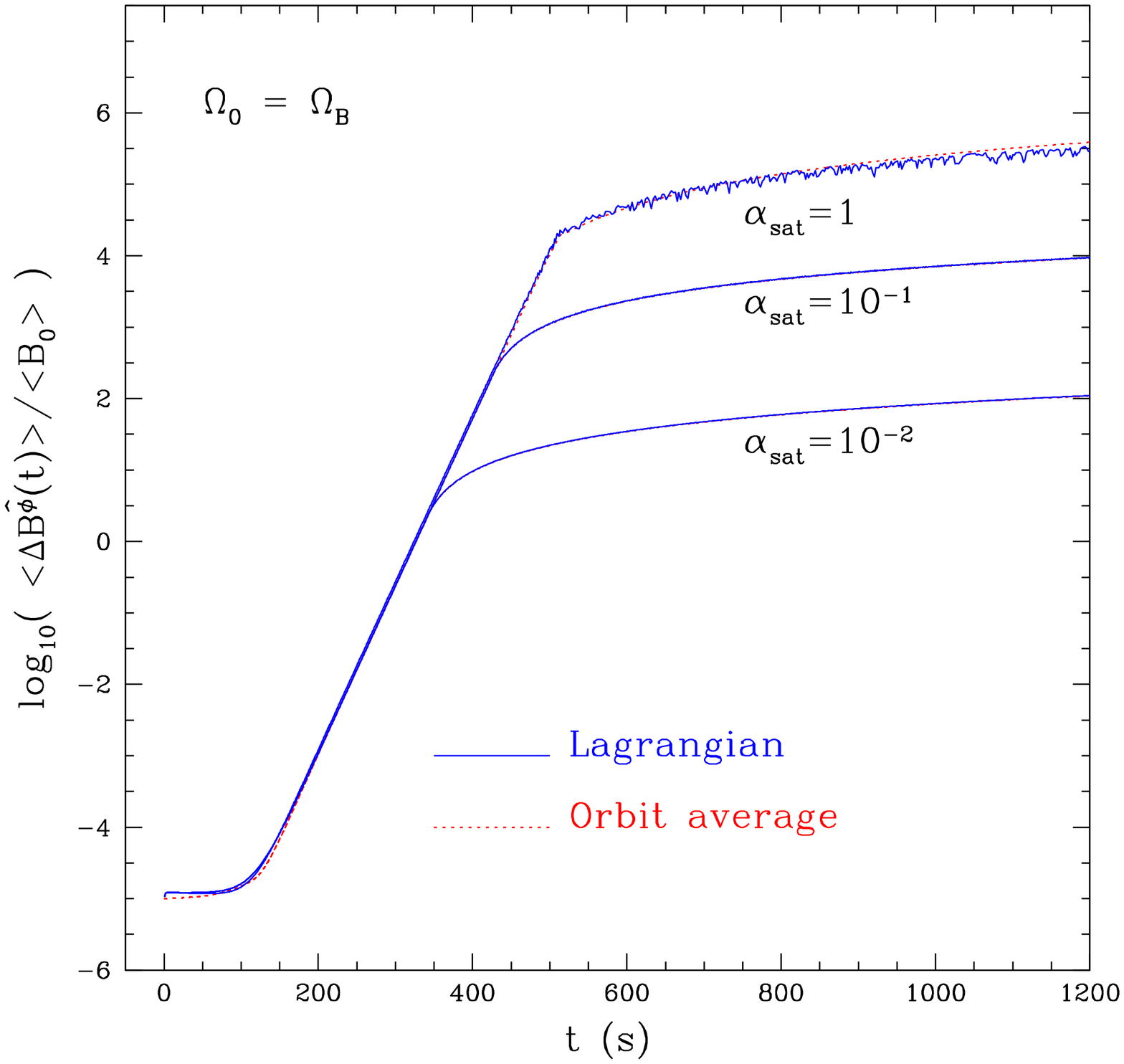,angle=0,width=8.5truecm,height=9.5truecm}
\hspace{0.25truecm}
\psfig{file=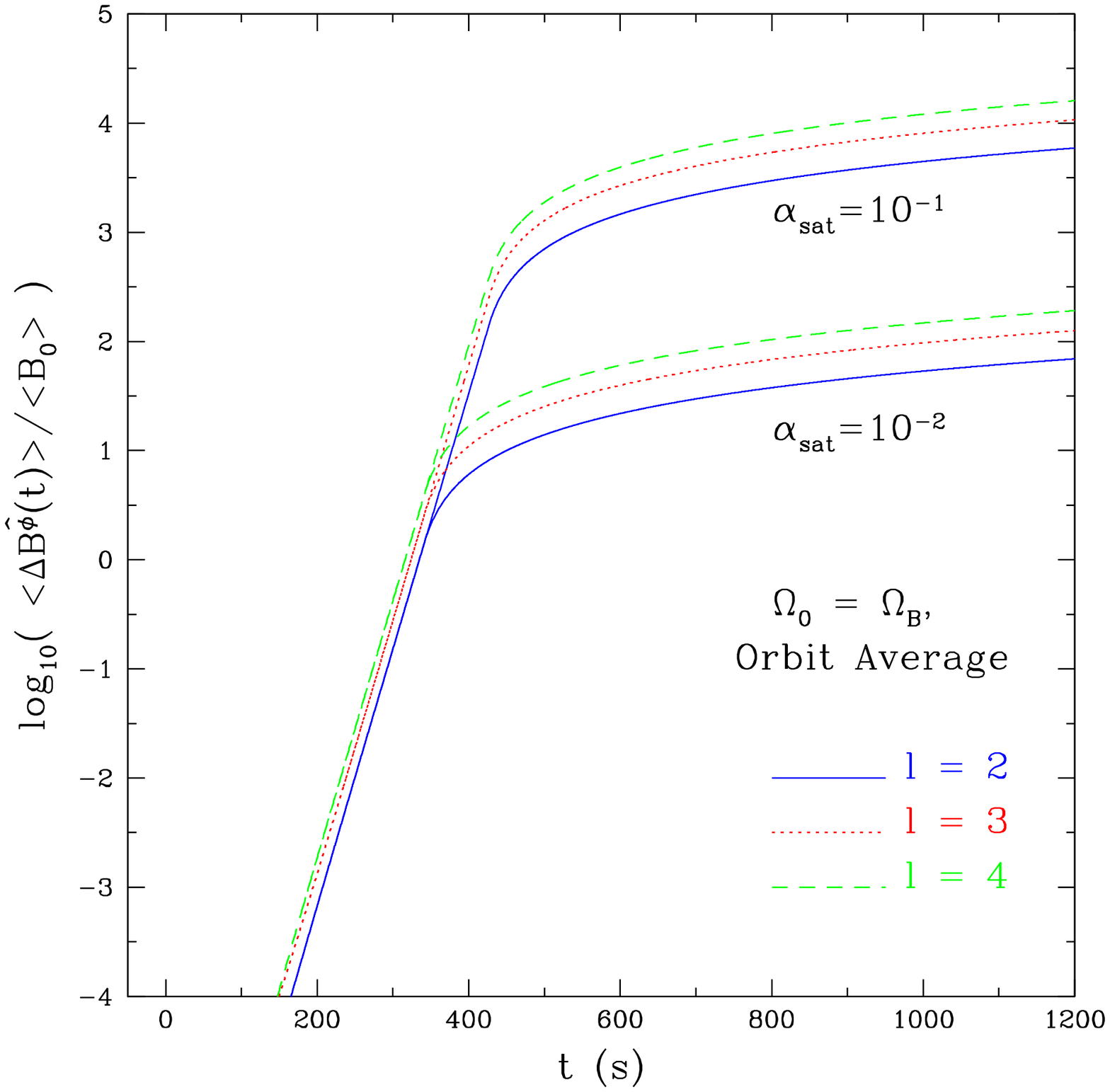,angle=0,width=8.5truecm,height=9.5truecm}
}
\caption[]{\label{fig3} Growth of the logarithm of
$\langle \Delta B^{\hat{\phi}}\rangle$ scaled to the
initial poloidal field. The left panel refers to an $\ell
= 2$ mode and a number of different values of
$\alpha_{\rm sat}$. The right panel shows the growth rate
for different $\ell$ modes and two different saturation
amplitudes.  The underlying star has an initial angular
velocity $\Omega_0$ equal to the break-up limit
$\Omega_{_B}$, and $\langle B_0\rangle=10^6$ G. Note that
in plotting the evolution of different mode numbers in
the right panel we have assumed they all have the same
growth time. This allows us to present them on a single
plot but one should bear in mind the higher-order modes
have considerably larger growth times~\cite{oetal98}.}
\end{center}
\end{figure}

	Indicating by $\langle \Delta B^{\hat {i}}
\rangle$ the $i$-th component of the volume average
secular magnetic field, we show in Fig.~\ref{fig3} the
time evolution of the averaged toroidal magnetic field
normalized to the average initial magnetic field. The
left panel of Fig.~\ref{fig3}, in particular, shows the
time evolution of the average toroidal field as computed
with the Lagrangian approach (continuous lines) and with
the orbit average approach (dotted lines). Different
lines refer to different values of the saturation
amplitude $\alpha_{\rm sat}$ and have been computed for
an $\ell = 2$ mode and a ``fiducial'' neutron star with
mass $M=1.4 {\rm M}_\odot$, radius $R = 12.5$ km, and
initial angular velocity $\Omega_0 $ equal to the
``break-up limit'' $\Omega_{_B} \equiv (2/3)\sqrt{\pi G
{\bar \rho}}$, with ${\bar \rho} = 3 M/(4 \pi R^3)$. The
value of the initial toroidal field served only to set a
lower limit to the vertical axis and was here set to be
\hbox{$10^{-6} \langle B_0\rangle$}. Each line in the
left panel shows an initial stage during which the
toroidal field grows exponentially and a subsequent one
during which the field has a growth which is a power-law
of time. As mentioned in the previous Section, the two
different regimes in the field growth reflect the fact
that there are two regimes in which secular kinematic
effects become apparent.

	The first regime is related to the exponential
growth of the mode amplitude and produces most of the new
magnetic field. During this phase, the secular effects
are extremely large [i.e. ${\cal O}(\alpha_0^2 \exp(t))$]
and the toroidal magnetic field is either produced or
amplified by the wrapping of the poloidal magnetic field
produced by the (mostly) toroidal secular velocity
field. The amplification is so large that the toroidal
magnetic field soon becomes comparable to and then larger
than the seed poloidal field. Larger values of the
saturation amplitude produce proportionally larger
amplifications and these can be so dramatic in the case
of $\alpha_{\rm sat}=1$, that the newly generated
magnetic field at mode saturation has become about four
orders of magnitude larger than any pre-existing magnetic
field. Note that a magnetic configuration with strong
toroidal magnetic fields is generally unstable and hoop
magnetic stresses in the radial direction will tend to
move magnetic field lines towards the
poles~\cite{b93}. Although the instability of this
configuration might lead to a number of interesting
astrophysical phenomena (see, for example,~\cite{s98}
where the buoyancy instability of this configuration is
used for a $\gamma$-ray burst model), we will here assume
that gas pressure gradients are always be able to
counteract these hoop stresses.

	The second regime is instead related to the
saturated stage of the transition, during which the mode
amplitude is constant in time. During this phase, the
secular effects are much smaller [i.e. ${\cal
O}(\alpha_{\rm sat}^2 t)$] and so is the field growth
rate, which is a power-law of time.

	The left panel of Fig.~\ref{fig3} also gives a
comparison between the Lagrangian and orbit-average
methods. Despite the differences between the two
approaches, the agreement is extremely good and
deviations between the two appear only for the late-time,
$\alpha_{\rm sat}=1$ growth rate, and at the beginning,
when the growth rate is extremely small. This difference
appears because of the difficulty of calculating the
particle trajectories accurately for such large values of
the mode amplitude and the inadequacy of the Lagrangian
method in this regime.

\begin{figure}[htb]
\begin{center}
\leavevmode
\hbox{
\hspace{0.125truecm}
\psfig{file=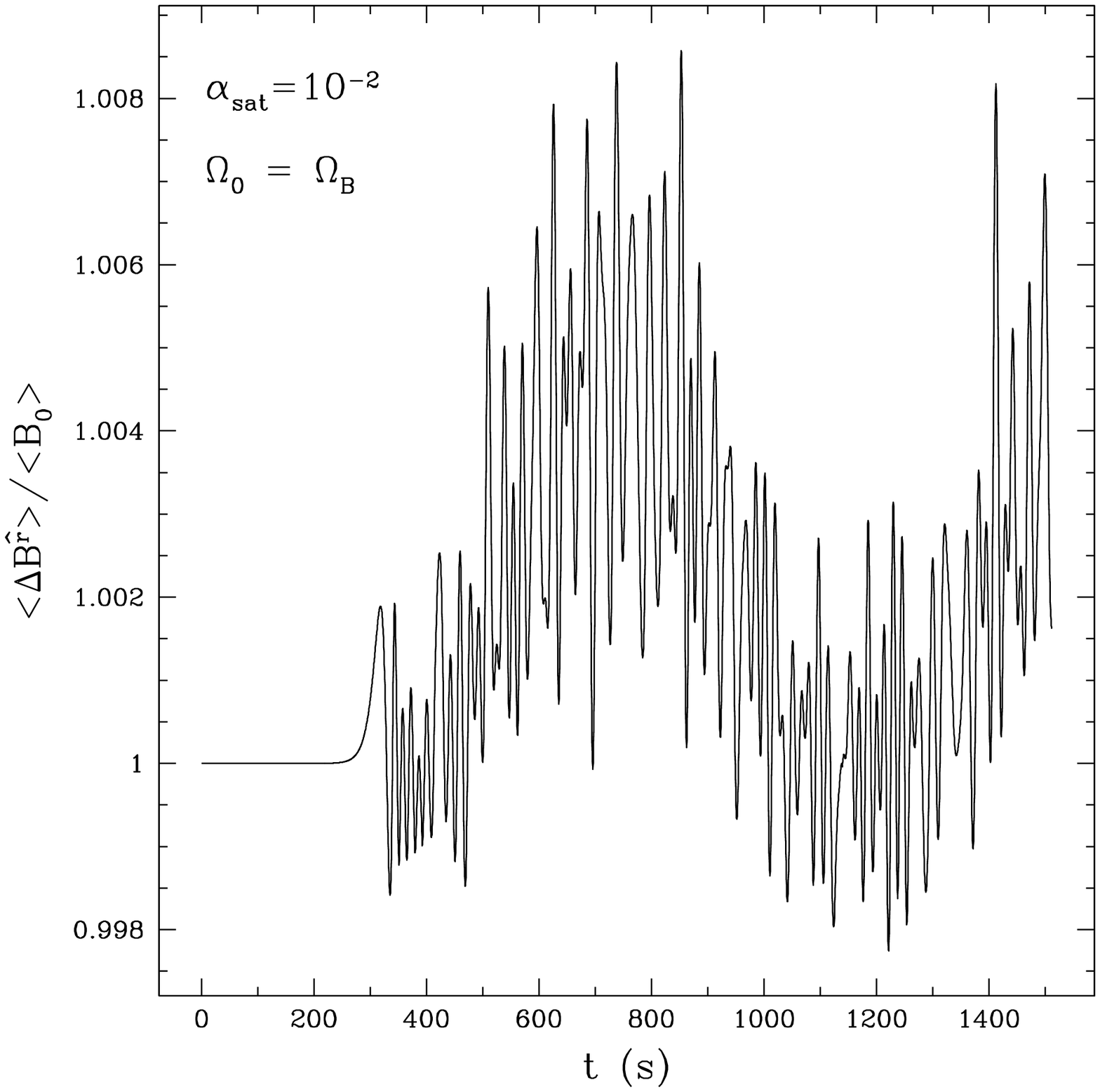,angle=0,width=8.5truecm,height=9.5truecm}
\hspace{0.25truecm}
\psfig{file=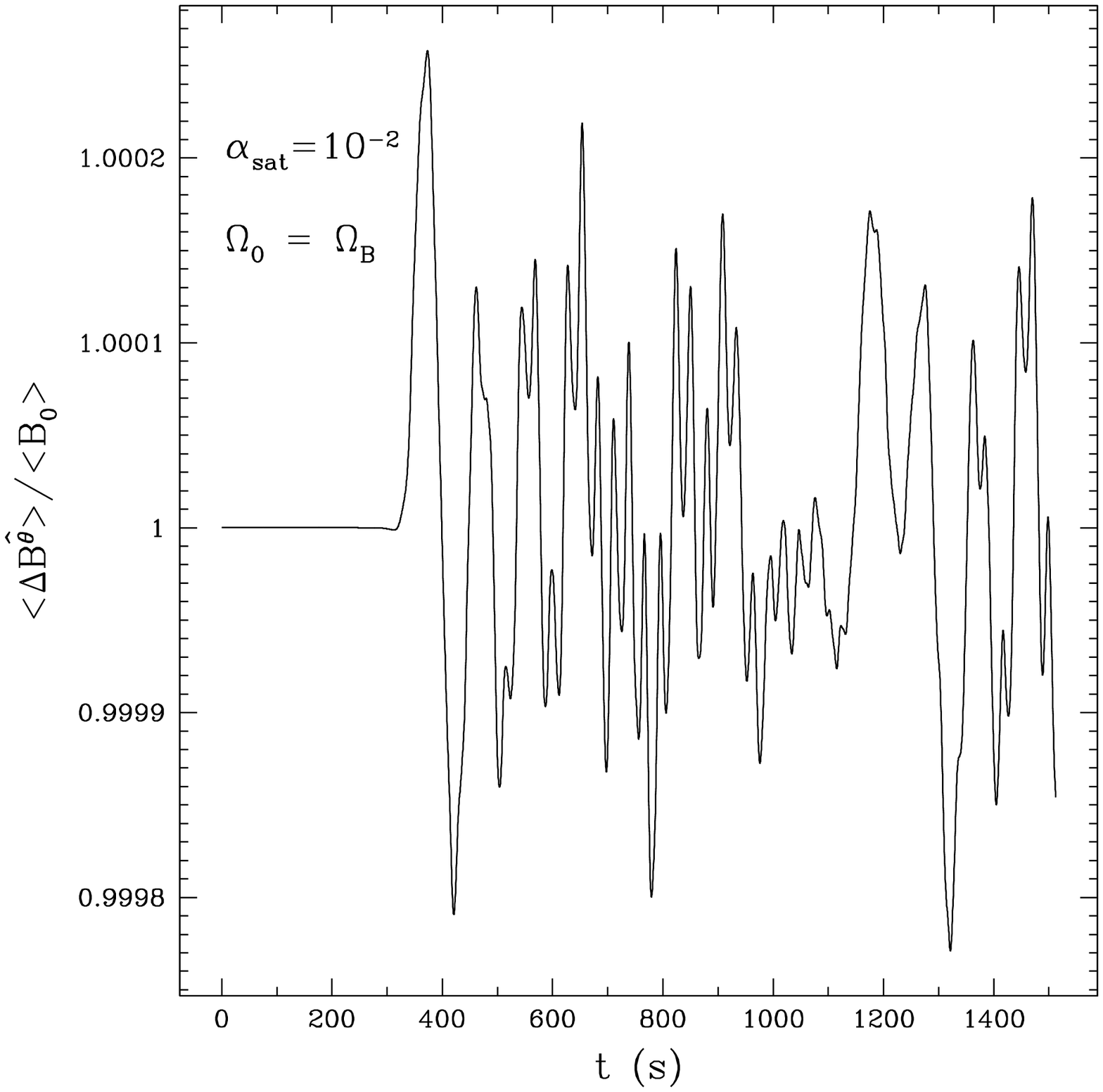,angle=0,width=8.5truecm,height=9.5truecm}
}
\caption[]{\label{fig4} Time evolution of $\langle \Delta
B^{\hat {r}}\rangle$ (left panel) and of $\langle \Delta
B^{\hat {\theta}}\rangle$ (right panel) scaled to the
initial field strength. The calculation has been done
with the orbit-average method for an $\ell = 2$ mode and
a saturation amplitude $\alpha_{\rm sat}=10^{-2}$.}
\end{center}
\end{figure}

	The right panel of Fig.~\ref{fig3} shows the same
quantity as the left one, but for different mode numbers
and two different saturation amplitudes. Note that, in
general, $r$ modes with different mode numbers saturate
on different timescales, with the high-order modes having
considerably longer growth times. However, in order to
present them all in the same plot, we have assumed in the
right panel of Fig.~\ref{fig3} that they all have the
same timescale. Interestingly, higher mode numbers would,
in principle, be more efficient in generating a toroidal
magnetic field, and thus in converting the energy of the
mode into magnetic energy, but modes with $\ell > 2$ are
also very unlikely to become unstable because of their
longer growth times.

	In Fig.~\ref{fig4} we show the corresponding
evolution of the poloidal magnetic field components for
an $\ell = 2$ mode and a saturation amplitude
$\alpha_{\rm sat}=10^{-2}$ as obtained through the orbit
average method. Note that neither of the two poloidal
magnetic field components shows a significant secular
growth and that they both oscillate around the initial
values. This is in agreement with Bragisnky's dynamo
model, whose velocity field has many analogies with the
secular velocity field set up by the $r$ modes, and which
predicts that the poloidal magnetic field should not be
subject to a dynamo action and have a zero time average
(see~\cite{m78} for a complete discussion of Braginsky's
dynamo). Fig.~\ref{fig4} basically states that a large
toroidal magnetic field will be the main outcome of the
secular kinematic effects produced during the $r$-mode
instability.

\begin{figure}[htb]
\begin{center}
\leavevmode
\hbox{
\hspace{0.125truecm}
\psfig{file=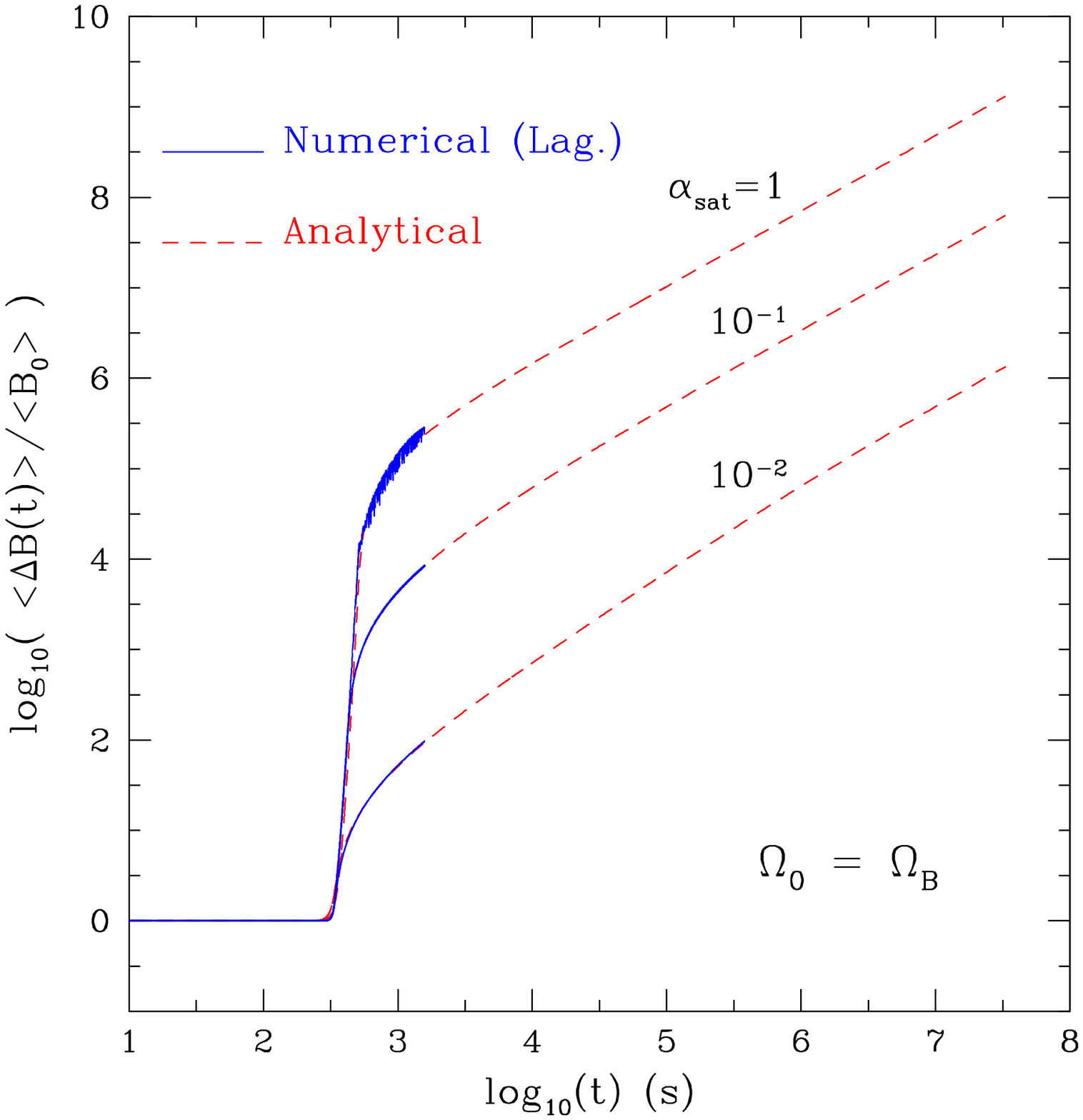,angle=0,width=8.5truecm,height=9.5truecm}
\hspace{0.25truecm}
\psfig{file=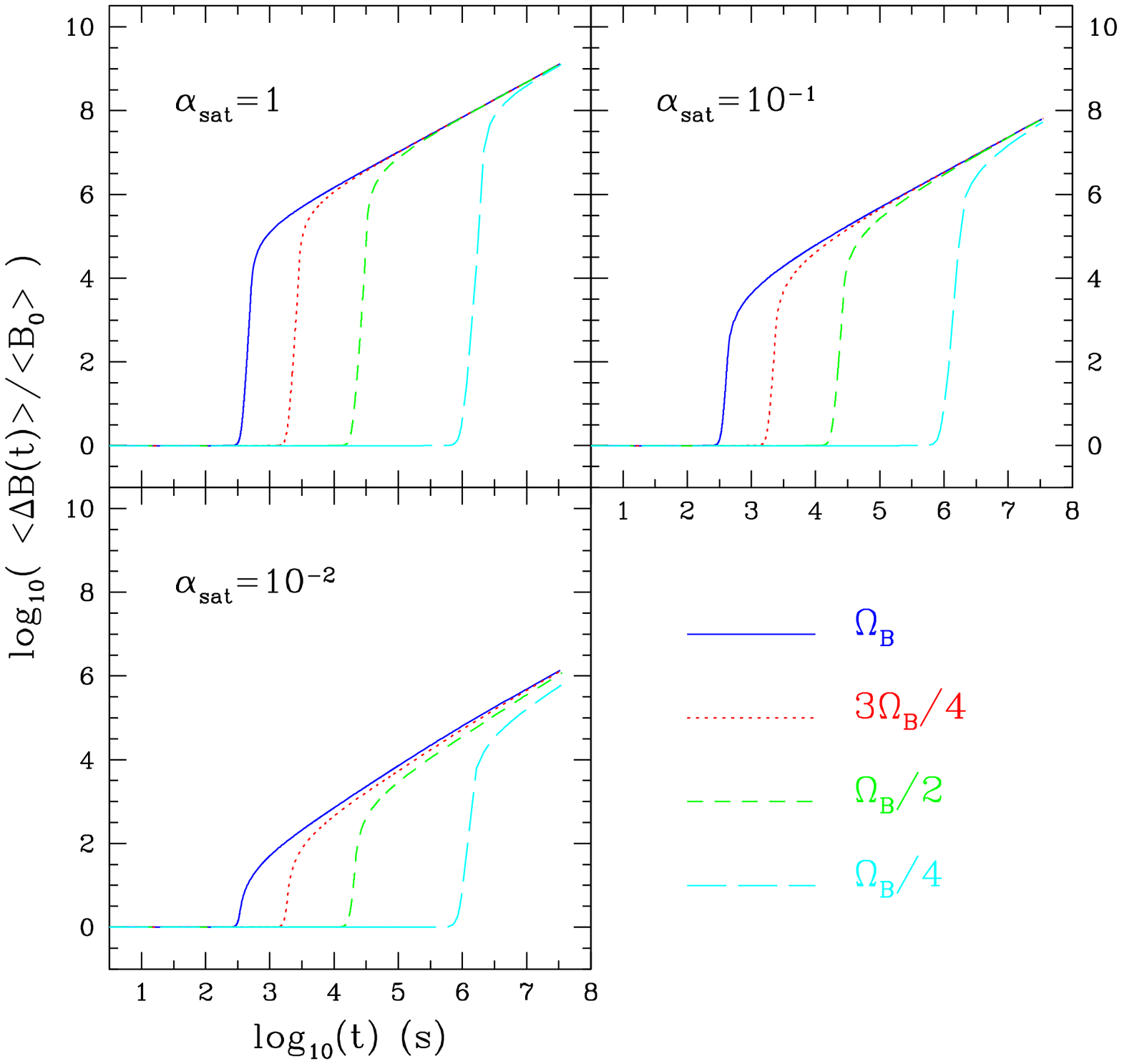,angle=0,width=8.5truecm,height=9.5truecm}
}
\caption[]{\label{fig5} The left panel shows the
long-time evolution of the logarithm of total magnetic
field $\langle \Delta B\rangle$ scaled to the initial
field for an $\ell = 2$ mode and different saturation
amplitudes. The star is set to have
$\Omega_0=\Omega_{_B}$. The right panel is the same as
the left one but shows the different evolutions resulting
from values of the initial angular velocity $\Omega_0 =
\frac{1}{4}\Omega_{_B},\ \Omega_0 =
\frac{1}{2}\Omega_{_B},\ \Omega_0 =
\frac{3}{4}\Omega_{_B},\ \Omega_0 = \Omega_{_B}$.  }
\end{center}
\end{figure}

	The numerical integration of the equations of
motion and the calculation of the magnetic fields
produced was usually stopped after 2000 s, corresponding
to $\sim 10^6$ \hbox{$r$-mode} oscillations. At that
stage and for our standard neutron star, the instability
has already reached saturation and the toroidal field
growth has entered a power-law growth phase which can be
described analytically (cf. Fig.~\ref{fig5}).

	Fig.~\ref{fig5} synthesizes the results presented
in the previous Figures and shows the time evolution of
the total magnetic field $\langle \Delta B\rangle$ scaled
to the initial magnetic field and over a timescale of one
year. The left panel, in particular, refers to an $\ell =
2$ mode, an initial angular velocity $\Omega_0 =
\Omega_{_B}$ and three different saturation
amplitudes. The panel also presents a comparison between
the fully numerical Lagrangian solution and the
analytical one based on equations (26) and (27) of paper
I. The agreement is remarkably good even for $\alpha_{\rm
sat} \simeq 1$ when we expect the expansion procedures in
powers of $\alpha$ to be rather poor.

	The right panel of Fig.~\ref{fig5} is the same as
the left one but also shows the effect of a smaller
initial angular velocity (i.e. $\Omega_0 =
\frac{1}{4}\Omega_{_B},\ \Omega_0 =
\frac{1}{2}\Omega_{_B},\ \Omega_0 =
\frac{3}{4}\Omega_{_B},\ \Omega_0 =
\Omega_{_B}$). Clearly, for stars rotating at slower
angular velocities, the instability and the field growth
set in at progressively later times. However, for any
given saturation amplitude, the final magnetic field
produced is rather insensitive of the rotation rate. This
is because the magnetic field is generated mostly during
the phase of exponential growth of the mode and as long
as the star rotates sufficiently fast for the mode to
grow unstable, the end result will remain roughly
unchanged.

\section{Impact of Magnetic Fields}
\label{iomf}

	In paper~I, as well as in the previous Sections
of this paper, we have described how $r$-mode
oscillations can produce secular drift velocity fields
which, in a highly conducting plasma such as neutron star
matter, can interact with pre-existing magnetic fields
producing new ones. We have also discussed the set of
equations that can be solved to calculate the magnitude
of the newly generated magnetic fields and have solved
them numerically to compute the values of the different
magnetic field components produced for a number of
different mode amplitudes and stellar angular
velocities. We have not yet discussed the impact of these
fields on the existence or growth of the $r$-mode
oscillations. As mentioned in Section~\ref{rmi_evol}, a
fully self-consistent treatment is beyond the scope of
this paper. However, important considerations can be made
if we treat the $r$-mode instability and the generation
of magnetic fields as evolving independently and with the
magnetic field not ``back-reacting'' on the kinematic
features of the \hbox{$r$-mode} oscillations. Under this
simplifying hypothesis, it is possible to calculate the
strength of the magnetic field necessary to {\it prevent}
the amplification of the first $r$-mode oscillation, or
{\it damp} the instability when this is free to
develop. Both of these aspects are discussed in the
following Sections~\ref{prev} and~\ref{damp}.

\section{Distortion of $r$ modes by the stellar magnetic field}
\label{prev}

	As the magnetic field strength increases, the
magnetic back-reaction forces generated by the $r$-mode
motions of the fluid become larger and larger. As the
magnetic field strength increases, the normal modes of
the star corresponding to the zero-magnetic-field $r$
modes differ in character more and more from the
zero-magnetic-field $r$ modes, taking on appreciable
hydromagnetic and hydromagnetic-inertial
properties. Eventually, the velocity perturbations
corresponding to these normal modes are significantly
different from the zero-magnetic-field $r$ modes to the
point of preventing gravitational radiation from
amplifying these distorted $r$-mode oscillations.

	Since the normal modes of the star are global in
character, the effects of a magnetic field cannot be
analyzed fully by a local analysis, such as can be done
by making a local comparison of forces. In contrast, a
mode-energy approach takes into account the global
properties of the modes and magnetic field and this is
what we will consider in the following. During an
oscillation the energy in the magnetic field will rise
and fall. If there is not enough energy in the mode to
supply the maximum magnetic energy increase required to
complete an oscillation, a ``full'' $r$-mode oscillation
will not occur (some other, smaller scale pulsations
might still occur). Stated differently, if the magnetic
field exceeds a critical value, $B_{\rm crit,p}$, the
magnetic stress that builds up during an oscillation will
be so large that it will halt the fluid motion involved
in the oscillation, i.e. the fluid momentum density will
be brought to zero. The condition that defines $B_{\rm
crit,p}$ is therefore
\begin{equation}
\label{bcp_def}
\delta E_{_{\rm M}} = \widetilde{E} \ ,
\end{equation}
where $\widetilde E$ the energy in the mode as measured
in the corotating frame of the equilibrium star
\begin{equation}
\label{e_inmode}
\widetilde{E}= \frac{1}{2} \frac{\alpha^2 \Omega^2}{R^2}
	\int^R_0 \rho\, r^6 dr \simeq 
	8.2 \times 10^{-3} \alpha^2 M \Omega^2 R^2 \ ,
\end{equation}
and 
$\delta E_{_{\rm M}}$ is the change in magnetic energy
density
\begin{equation}
\label{deltaem}
\delta E_{_{\rm M}} \equiv \frac{1}{8\pi}
	\int_{V_{\infty}}\delta B^2 {\;d^3{\bf x}}\ ,
\end{equation}
with $\delta B^2 = (\delta B^{\hat {p}})^2 + (\delta
B^{\hat {\phi}})^2$. Note that the fractional change in
the mode energy produced by gravitational wave emission
during a single oscillation period is
$4\pi/\omega|\tau_{_{\rm GW}}|$. This is always $\ll 1$ and
therefore can be neglected in this comparison.

	The expression for $\delta E_{_{\rm M}}$ varies
according to whether the neutron star core is composed of
a normal neutron fluid (N) or is superconducting
(SC). More particularly, we denote as $\delta E^{\rm
N}_{_{\rm M}}$ and $B^{\rm N}_{\rm crit,p}$ the magnetic
energy and the critical, volume averaged, magnetic field
for a normal core. Similarly, we denote as $\delta
E^{^{\rm SC}}_{_{\rm M}}$ and $B^{^{\rm SC}}_{\rm
crit,p}$ the corresponding quantities for a
superconducting core. The two cases will be considered
separately below.

\subsection{When the stellar core is normal}

	The induction equation can be integrated in time
to estimate the variations in the magnetic field produced
by the perturbation velocity field as 
\begin{eqnarray}
\label{ddbs}
&&\delta B^{\hat {\theta}} \simeq B^{\hat {\phi}}_0 
	\int {\dot \theta}(t') dt' 
	\simeq B^{\hat {\phi}}_0 \int \frac{\delta
	v(t')}{r} dt' \simeq \delta B^{\hat {p}}\ , 
\\
&&\delta B^{\hat {\phi}} \simeq B^{\hat {\theta}}_0 
	\int {\dot \phi}(t') dt' 
	\simeq B^{\hat {\theta}}_0 \int \frac{\delta
	v(t')}{r} dt' \ , 
\end{eqnarray}
where we have set ${\dot \theta} \simeq {\dot \phi}
\simeq \delta v/r$. Because of the periodic variation of
the magnetic energy during an oscillation, expression
(\ref{deltaem}) should be averaged over half of an
oscillation period, so that the magnetic energy produced
is
\begin{equation}
\label{em_qrtr}
\delta E^{\rm N}_{_{\rm M}} = \left(\frac{9\pi}{32}\right)\frac{1}{\Omega^2}
	\int_{V_*}\left[B_0 \frac{\delta v}{r}\right]^2 
	{\;d^3{\bf x}} \ .
\end{equation}
In deriving (\ref{em_qrtr}) we have not included
contributions to $\delta E_{_{\rm M}}$ from changes in
the magnetic field outside the star which are heavily
suppressed by the high electric conductivity of the
crust. Also, we have written the time average of the
velocity perturbation as
\begin{equation}
\int^{P/2}_0 \delta v(t') dt' \simeq \left(\frac{3\pi}{2}\right) 
	\frac{\delta v}{\Omega} \ .
\end{equation}
Note that both energies (\ref{e_inmode}) and
(\ref{em_qrtr}) depend quadratically on the velocity
perturbation $\delta v$ and therefore on $\alpha$. As a
result, the critical fields are not not sensitive to the
amplitude development of the mode. The volume integral
can be easily performed after assuming an initial dipolar
magnetic field: $B_0 = B_{\rm d} (R/r)^3 \Psi(\theta)$,
where \hbox{$\Psi(\theta)\equiv 4 \cos^2\theta +
\sin^2\theta$}. Indicating with $|{\mathbf Y}^B_{2 2}
(\theta,\phi)|$ the modulus of the vector spherical
harmonic, the magnetic energy produced over half of a
period is then
\begin{equation}
\label{enm}
\delta E^{^{\rm N}}_{_{\rm M}} = 
	\left(\frac{9\pi}{32}\right) B^2_{\rm d} R^4 \alpha^2 
	\int^{2\pi}_0 d\phi \int^{\pi}_0 
	|{\mathbf Y}^B_{2 2}(\theta,\phi)| 
	\Psi(\theta)\sin\theta d\theta 
	\int^R_{pR} r^{-2} dr = 
	\left[\frac{3\pi(1-p)}{32p}\right] \Lambda \alpha^2 
	B^2_{\rm d} R^3 \ ,
\end{equation}
where all of the angular dependence is contained in
$\Lambda$
\begin{equation}
\label{lambda}
\Lambda \equiv
	\int^{2\pi}_0 d\phi \int^{\pi}_0 |{\mathbf Y}^B_{22}
	(\theta,\phi)| \Psi(\theta)\sin\theta d\theta
	= {\mathcal O}(1) \ .
\end{equation}
It is worth noticing that we can here neglect the
distortion of the magnetic field near the crust-core
boundary during an $r$-mode oscillation. The reason for
this is that, on the timescales relevant for the
$r$ modes and regardless of whether the core is
superconducting, the magnetic field lines (and flux
tubes) are pinned in place at the crust-core boundary by
the high conductivity of the crust.

	Using (\ref{enm}) with $p=0.5$ and
(\ref{lambda}), we can now write the condition for
$r$-mode prevention (\ref{bcp_def}) in terms of a
critical field given by
\begin{equation}
\label{bncp}
B^{^{\rm N}}_{\rm crit,p} \simeq 2.5 \times 10^{16}
	\left(\frac{\Omega}{\Omega^*_B}\right)
	\left(M_{1.4}\right)^{1/2} 
	\left(R_{12.5}\right)^{-1/2}\ {\rm G}\ , 
\end{equation}
where $M_{1.4} \equiv M/1.4\,M_\odot$, $R_{12.5} \equiv
R/12.5\,{\rm km}$, and $\Omega^*_B \equiv 5.62 \times
10^3$~rad~s$^{-1}$.

\subsection{When the stellar core is superconducting}

	When, as a result of the star's cooling, the core
temperature falls below a critical value, $T_c \sim
3\times 10^{9}$~K, the core is expected to become a
Type~II superconductor~\cite{l91}. In such a
superconducting core, the magnetic field is confined to
flux tubes and the total energy of the flux tubes can be
computed in terms of the total number $N$ of flux tubes
in the core and of the mean total energy of a flux tube
\begin{equation}
\langle e \rangle = \frac{\phi_0}{8 \pi} 
	\langle H_f \ell_f \rangle \ ,
\end{equation}
with $\ell_f$ being the average length of a flux tube,
$H_f$ the magnetic field inside it and $\phi_0 = h
c/(2e)\simeq 2\times10^{-7}$~G cm$^2$ the flux
quantum. The total magnetic energy of the flux tubes of
the unperturbed star is then
\begin{equation}
E^{^{\rm SC}}_{_{\rm M}} = \frac{\phi_{\rm tot}}{8\pi} 
	\langle H_{f} \ell_f \rangle \ ,
\end{equation}
where $\phi_{\rm tot} \approx \pi R^2 B_{\rm d}$ is the
total magnetic flux. The $r$-mode oscillations will shear
and stretch the flux tubes, increasing their length by
${\delta\ell}$, and thus changing their energy content by
an amount $\delta E^{^{\rm SC}}_{_{\rm M}} = ({\phi_{\rm
tot}}/{8\pi}) H_{f} \delta \ell_f$ over half a
period. Estimating the volume-averaged variation in the
flux tube length as
\begin{equation}
\delta \ell \simeq \frac{9\pi^2}{56} \alpha^2 R \ , 
\end{equation}
we then obtain
\begin{equation}
\label{descm}
\delta E^{^{\rm SC}}_{_{\rm M}} \simeq \frac{27 \pi^2}{448} \alpha^2
	H_f B_{\rm d} R^3 
	\simeq 4.5 \times 10^{34} \alpha^2 
	\left(\frac{H_{_{C1}}}{4.4\times 10^{15}} \right)
	B_{\rm d} R^3_{12.5} \ {\rm ergs} \ ,
\end{equation}
where we have taken $H_f \approx H_{_{C1}}$, the lower
critical field~\cite{dg66}.

	Note that expression (\ref{descm}) is only valid
when the spacing between the flux tubes is much larger
than the London penetration length, i.e. for $\langle B
\rangle \ll H_{_{C1}}$. However, as $\langle B \rangle$
rises above $H_{_{C1}}$ and tends to the upper critical
field $H_{_{C2}}\approx 10^{17}$~G, $H_f$ also rises and
tends to $H_{_{C2}}$, with the flux tubes crowding one
another. When $\langle B \rangle= H_{_{C2}}$, the normal
cores of the flux tubes touch, the superconducting state
is destroyed~\cite{dg66} and the variation in the
magnetic energy is then again given by expression
(\ref{enm}). Combining (\ref{descm}) with (\ref{enm}), we
can now write the condition for $r$-mode prevention in
terms of the critical field for a superconducting core
\begin{equation}
\label{bsccp}
B^{^{\rm SC}}_{\rm crit,p} \simeq 9.7 \times 10^{16}
	\left(\frac{\Omega}{\Omega^*_B}\right)^2
	M_{1.4}	\left(R_{12.5}\right)^{-1} 
	\left(\frac{H_{_{C1}}}{10^{16}\ {\rm G}}\right)^{-1} 
	\ {\rm G} \ .
\end{equation}
As for the critical field for a normal core, expression
(\ref{bsccp}) is insensitive to the values of the mode
saturation amplitude, underlining that if $B_0 > B_{\rm
crit,p}$ even very small amplitude fluctuations of
$r$-mode character cannot occur. On the other hand, the
critical fields are sensitive to the star's spin rate; in
particular, if $\Omega$ is $\approx \Omega^*_B$, the
magnetic field needed to prevent oscillation of the
$\ell=2$ mode is $\sim 10^{16}\,{\rm G}$ and hence
$B_{{\rm crit},p} \sim B^{N}_{{\rm crit},p}$, whether or
not the core is superconducting. If instead $\Omega \ll
\Omega^*_{B}$, then $B_{{\rm crit},p}^{SC} \approx
10^{13}\, (\Omega/0.01\,\Omega_B^*)^2
H_{C1,16}^{-1}M_{1.4} R_{12.5}^{-1}\,$G, where $H_{C1,16}
\equiv H_{C1}/10^{16}\,{\rm G}$.

	The critical fields (\ref{bncp}) and
(\ref{bsccp}) are several orders of magnitude larger than
the typical magnetic fields of young neutron stars that
observations indicate to be in the range $10^{11} -
10^{13}$~G. This suggests that the distortion of the
first $r$-mode oscillations by magnetic fields is rather
unlikely and that the $r$-mode instability in newly born
neutron stars (or any compact object with analogous
properties) is, in general, free to develop. The onset of
the instability, however, will also lead to the
production of very intense magnetic fields which could
ultimately lead to the damping of the oscillations, as
discussed below.

\section{Damping of $r$ modes by the stellar magnetic field}
\label{damp}

	The damping of oscillations by the growing
magnetic field reflects the question of whether the
energy required to continue the oscillation for the next
time step $\Delta t$ exceeds the energy that can be
pumped into the mode in the time $\Delta t$ by the
emission of gravitational radiation. 

\subsection{When the stellar core is normal}

	A critical point in the balance between the
energy spent in producing magnetic field and the energy
in the mode provided by the emission of gravitational
waves is reached when the two rates are equal
\begin{equation}
\label{equal_rates} 
\frac{dE_{_{\rm M}}}{dt} = 
\left(\frac{d{\widetilde E}}{dt}\right)_{_{\rm GW}} \ , 
\end{equation}
where 
\begin{equation}
\label{de_explicit}
\left( {d\widetilde{E}\over dt} \right)_{_{\rm GW}} =
	- {32\pi G \over c^{7}}
	\left[ {4\over 3(5)!!} \right]^2
	\frac{\alpha^2 \Omega^4 \omega^{4}}{R^{2}}
	\left( \int_0^R\rho\,r^{6} dr \right)^2 \ ,
\end{equation}
is the rate of energy transferred to an $\ell=2$
mode~\cite{oetal98}. After the equality
(\ref{equal_rates}) is reached, ${dE_{_{\rm M}}}/{dt} >
(d{\widetilde E}/{dt})_{_{\rm GW}}$. The reason for this
is that the toroidal magnetic field, and hence
${dE_{_{\rm M}}}/{dt}$, will continue to grow whereas
$(d{\widetilde E}/{dt})_{_{\rm GW}}$ remains the same or
decreases as gravitational radiation causes the star to
spin down. The only source of energy to feed $E_{_{\rm
M}}$ is thus the energy ${\widetilde E}$ of the mode
which damps on a timescale
\begin{equation}
\tau_{_{\rm M}} = \frac{\widetilde E}{({dE_{_{\rm M}}}/{dt})} \ .
\end{equation}
	
	A measure of the relative importance of the two
energy rates can be obtained through considering their
ratio
\begin{equation}
\label{gt}
g(t) \equiv \frac{{dE_{_{\rm M}}}/{dt}}
	{(d{\widetilde E}/{dt})_{_{\rm GW}}} \ ,
\end{equation}
where, in the case of normal core, the rate of production
of magnetic energy takes the form
\begin{equation}
\label{edot_nc}
\frac{d E^{^{\rm N}}_m(t)}{dt} = 
	\left[\frac{4(1-p)}{9\pi p}\right] B^2_{_S} R^3
	\Lambda' \alpha^2(t)\Omega(t) 
	\int^t_0 \alpha^2(t')\Omega(t')dt' \ ,
\end{equation}
with now $\Lambda' \equiv \int^{2\pi}_0 d\phi
\int^{\pi}_0 (\kappa_2(\theta))^2 |{\mathbf
Y}^B_{22}(\theta,\phi)| \Psi(\theta)\sin\theta d\theta$,
which is again ${\mathcal O}(1)$.

\begin{figure}[htb]
\begin{center}
\leavevmode
\hbox{
\hspace{0.125truecm}
\psfig{file=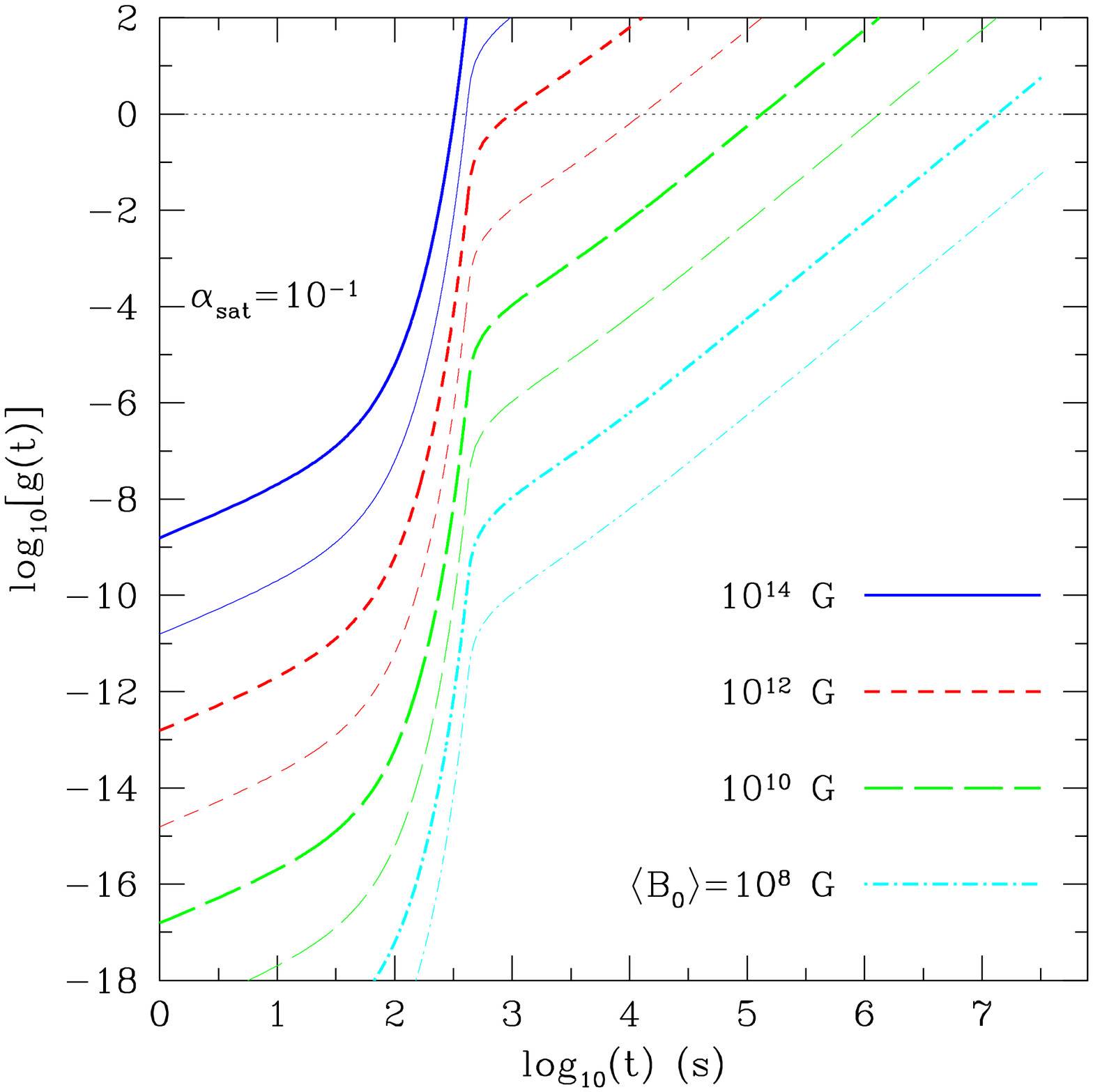,angle=0,width=8.5truecm,height=9.5truecm}
\hspace{0.0truecm}
\psfig{file=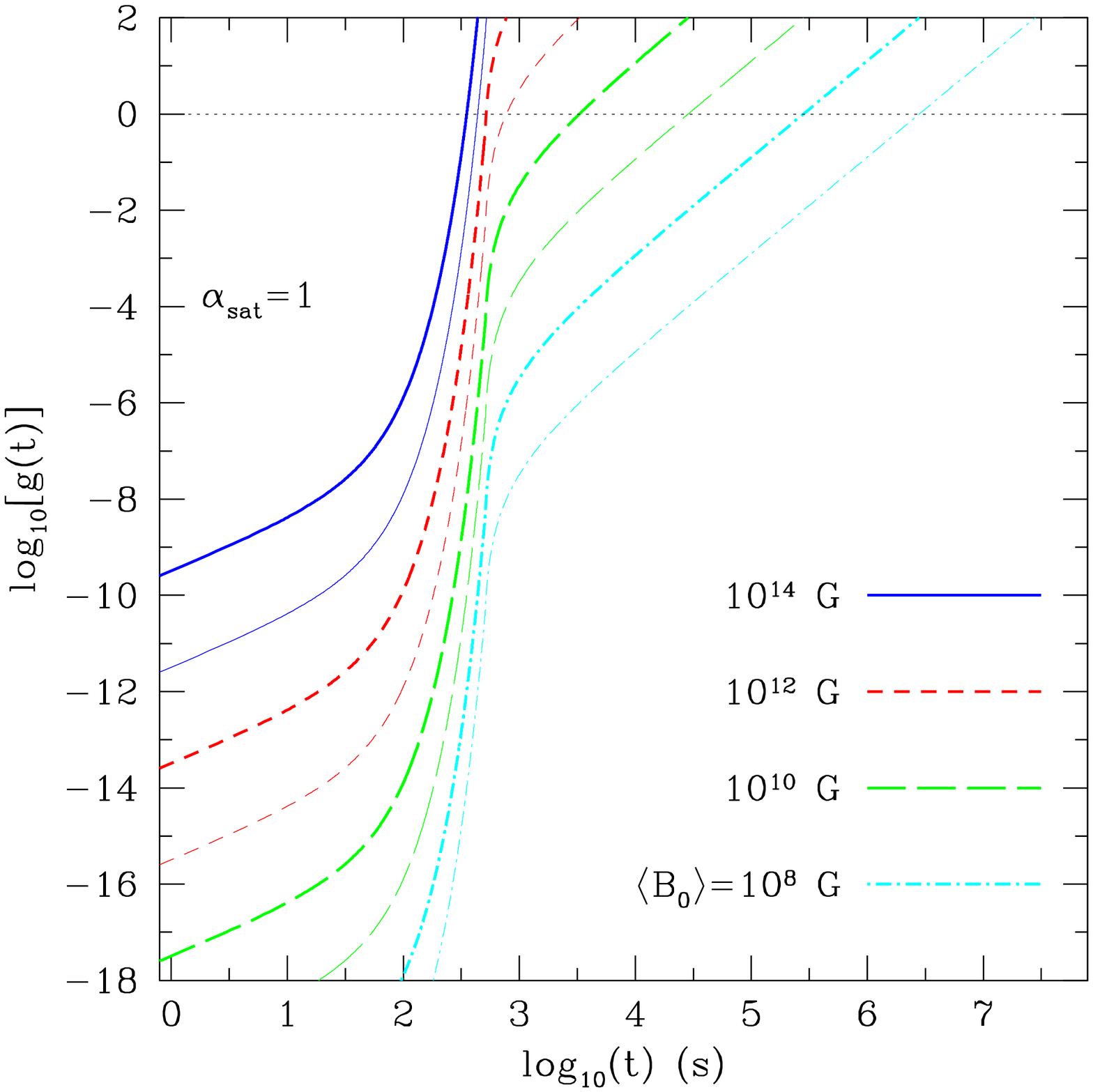,angle=0,width=8.5truecm,height=9.5truecm}}
\caption[]{\label{fig6} Time evolution of the ratio
$g(t)$. The calculation refers to our ``fiducial''
neutron star initially rotating at the break-up limit
$\Omega^*_B$. The left and right panels show the
evolution for different values of the saturation
amplitude. Lines of different type refer to different
values of the initial magnetic field $\langle B_0
\rangle$. Nearby thick and thin lines of the same type
are used to compare the magnetic field evolution produced
by the secular velocity field calculated in paper~I
[cf. Eq. (7) of paper~I] and a velocity field which is
just a tenth of it.}
\end{center}
\end{figure}

	Fig.~\ref{fig6} shows the numerical computation
of the ratio (\ref{gt}) for a ``fiducial'' neutron star
initially rotating at the break-up limit
$\Omega^*_B$. The rate of energy loss to gravitational
radiation has been computed for an $\ell = 2$ mode and a
polytropic equation of state with $\Gamma=2$. The left
panel shows the time evolution of $g(t)$ for saturation
amplitudes $\alpha_{\rm sat}=0.1$ and four different
values of the initial magnetic field $\langle B_0
\rangle$ which are indicated with different types of
thick lines. The right panel is the same as the left one
except for having $\alpha_{\rm sat}=1$. For each curve in
Fig.~\ref{fig6} it is straightforward to distinguish the
initial phase of exponential growth of the magnetic field
(and of the magnetic energy production rate) from the
subsequent phase of power-law growth. It also appears
evident that, for a given saturation amplitude, different
values of $\langle B_0 \rangle$ have only the effect of
moving the curves in the vertical direction: the smaller
the initial field, the longer it will take to reach a
sufficiently strong magnetic field necessary for
suppression.

	As discussed in paper~I, we expect the drift of
fluid elements given by the velocity field ${\mathbf
v}_{\rm d}$ [cf. Eq. (7) of paper~I] to be qualitatively
correct to ${\cal O}(\alpha^2)$. However, within our
approach we cannot exclude the possibility that the
actual drift velocity is smaller (or larger) than the
estimated one. To assess the variations brought about by
a smaller secular velocity field, we use the thin lines
in Fig.~\ref{fig6} that are near to the thick lines of
the same type to show the magnetic field evolution which
has been calculated with the secular velocity field being
one tenth of that computed in paper I. Because of the
exponential growth in the mode amplitude and in the
toroidal magnetic field, the overall differences produced
are rather small.
	
	When read in terms of the times at which
suppression of the oscillations begins, Fig.~\ref{fig6}
indicates that for an initial magnetic field $\langle B_0
\rangle = 10^{10}-10^{12}$~G, the instability starts to
be suppressed after a time between 2 days and 15 minutes
for a saturation amplitude $\alpha_{\rm sat}=0.1$ (see
also Table I). When this happens the total magnetic
fields produced are in the range $10^{14}-10^{15}$ G.
The condition for oscillation suppression
(\ref{equal_rates}) can also be used to calculate
$\Omega_{\rm crit}$: the critical spin rate above which
$r$~modes are excited and below which they are
damped. Values of the critical angular velocity for a
normal core normalized to the break-up angular velocity
$\Omega^*_{B}$ are indicated in Table I. The Table
reports the critical spin rates as a function of the
initial magnetic field strengths for a number of
different saturation amplitudes. For each pair of these
parameters, we also indicate in brackets the time (in
seconds) $t_{\rm crit}$ at which the condition
(\ref{equal_rates}) is reached. As we will discuss in
Section~\ref{gwd}, the knowledge of the critical angular
velocity is in fact very important as it provides the
final (and maximum) gravitational wave frequency at which
the last gravitational radiation is emitted. Together
with the initial star's spin frequency, the maximum
frequency determines the domain over which a
signal-to-noise ratio can be calculated
[cf. Eq.~(\ref{s_o_n})].

	It is worth underlining that even when the
$r$-mode instability is rapidly suppressed by generation
of intense magnetic fields, there is sufficient time for
the star to loose a good fraction of its rotational
energy to gravitational waves. This is because the
loss-rate of angular momentum to gravitational waves is a
steep function of the star's angular velocity
[cf. Eq.~(\ref{odot_bs})] and is large mostly soon after
the amplitude saturation. Therefore, for initial magnetic
fields of the order of $10^{10}$ G, most of the
astrophysical considerations about the importance of the
$r$-mode instability in spinning down newly-born neutron
stars~\cite{ak00} are basically unchanged.
\vbox{
\begin{table}[htb]
\begin{tabular}{|c|cccc|}
         &        &         &       &     			\\
$\Omega^{^{\rm N}}_{\rm crit}/\Omega^*_{B}, \ \ 
(t_{\rm crit}$ in s) 
	 & $\langle B_0 \rangle = 10^{8}\  $G 	 
	 & $\langle B_0 \rangle = 10^{10}\ $G 
	 & $\langle B_0 \rangle = 10^{12}\ $G 	 
	 & $\langle B_0 \rangle = 10^{14}\ $G 			\\ 
         &        &         &       &     			\\
\tableline 
 			    	&        &        &        &   	\\
$\alpha_{\rm sat}= 1.0~~~$  	& 0.318 \ \ ($2.90\times10^5$)
                            	& 0.467 \ \ ($3.34\times10^3$)
			    	& 0.948 \ \ ($5.24\times10^2$)
			    	& 0.999 \ \ ($3.52\times10^2$)\ \\  
$\alpha_{\rm sat}= 10^{-1}$ 	& --     
				& 0.535 \ \ ($1.37\times10^5$)
				& 0.975 \ \ ($9.75\times10^2$)
				& 0.999 \ \ ($3.23\times10^2$)\ \\ 
$\alpha_{\rm sat}= 10^{-2}$ 	& --     
				& --     
				& 0.974 \ \ ($5.62\times10^4$)
				& 0.999 \ \ ($3.23\times10^2$)\ \\ 

		            &        &        &        &     	\\
\end{tabular}
\label{table1}
\bigskip
\caption[]{Normalized critical angular velocities
[i.e. final values of the star's angular velocity when
$g(t)=1$, normalized to the initial break-up angular
velocity $\Omega^*_B$] for different values of the
magnetic field $\langle B_0 \rangle$ and saturation
amplitude $\alpha_{\rm sat}$. Values not reported refer
to times longer than one year, for which the instability
is suppressed by viscosity rather than by the production
of magnetic field. The numbers in brackets indicate the
time (in seconds) after which the condition for
suppression is reached.}
\end{table}}

\subsection{When the stellar core is superconducting}

	In the case of a superconducting core, the
calculation of the damping of the $r$ mode oscillations
is slightly more complicated (and uncertain) because
additional considerations need to be made about the time
variations of the total magnetic flux $\phi_{\rm tot}$
and about the magnetic field components that contribute
mostly to the flux. Once $B^{\hat\phi}$ exceeds $B^{\hat
p}$, the magnetic energy rate can be expressed as
\begin{equation}
\label{edot_scc}
\frac{d E^{^{\rm SC}}_{_{\rm M}}(t)}{dt} \simeq
	\frac{\kappa_2(\theta)}{60}
	H_{_{C1}} B^{\hat p} R^3 \alpha^2(t)\Omega(t) \ .
\end{equation}
Hence, for a superconducting core, the actual critical
spin rate is likely to be
\begin{equation}
\Omega^{^{\rm SC}}_{\rm crit} \simeq 0.7\,\Omega_B^*\, 
	\left(\frac{B^{\hat p}}{10^{12}{\rm G}}\right)^{1/7}
	\left(\frac{H_f}{10^{16}{\rm G}}\right)^{1/7} 
	M_{1.4}^{-2/7} R_{12.5}^{-3/7} \ ,
\end{equation}
which is again independent of the mode amplitude.

\section{Gravitational wave detectability}
\label{gwd}

	The relevance of the gravitational radiation
emitted by $r$-modes oscillations has been investigated
by a number of authors who have considered them both as
single sources~\cite{oetal98,bc00,fhh01} and as a
cosmological population producing a stochastic
background~\cite{fms99}. Their estimates indicate that
sufficiently close sources might be detectable by LIGO
and VIRGO when they reach their ``enhanced'' level of
sensitivity or that two nearby interferometers with
``advanced'' sensitivities might be used for the
detection.

	Using the results obtained in the previous
Section and bearing in mind the simplifying assumptions
under which they have been derived, we now re-evaluate
the possibility of detecting gravitational waves from
$r$ modes in magnetic neutron stars. All of the results
discussed below refer to a ``fiducial'' neutron star with
a normal core. For an $\ell = 2$ mode, the rate of
angular momentum loss due to the emission of
gravitational waves can be written as~\cite{oetal98}
\begin{equation}
{dJ_c\over dt} = \frac{D^2}{2} \omega
	\langle\langle h_+^2 +h_\times^2\rangle\rangle \ ,
\end{equation}
where $h_+$ and $h_\times$ are the strain amplitudes of
the two polarizations of gravitational waves, $D$ is the
distance to the source, and $\langle\langle \ldots
\rangle\rangle$ denotes the average over the orientation
of the source and its location on the sky. Here $J_c= - 3
\tilde{J} M R^2 \Omega_s \alpha^2$ is the canonical
angular momentum~\cite{fs78} (see also~\cite{oetal98})
and $\tilde{J}= \int_0^R \rho\, r^{6}dr/(MR^4)$.

\begin{figure}[htb]
\begin{center}
\leavevmode
\psfig{file=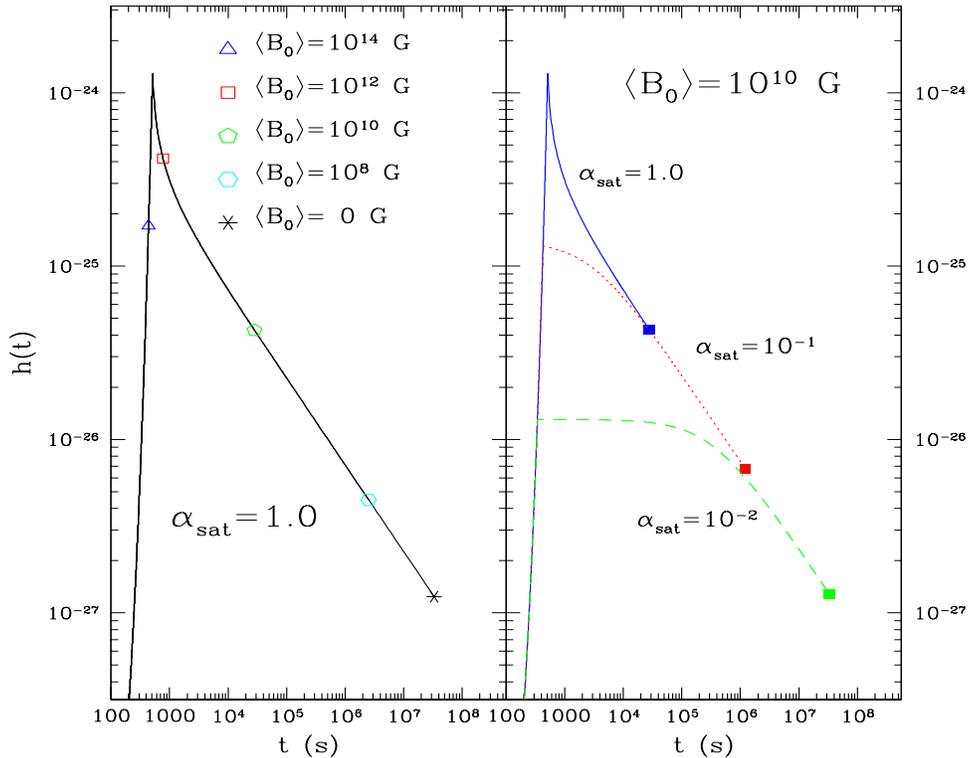,angle=0,width=13.0truecm,height=11.0truecm}
\caption[]{\label{fig7} Time evolution of the averaged
gravitational wave amplitude. The left panel is for a
saturation amplitude $\alpha_{\rm sat}=1$ and shows the
different behaviour with different initial values of the
magnetic field. The evolutionary tracks terminate when
$g(t)=1$ and different final points are marked with
different open symbols. The right panel refers to an
initial field of $\langle B_0 \rangle=10^{12}$ G and
shows the influence of the saturation amplitude on the
evolution of the wave amplitude. A filled square
indicates the time at which $g(t)=1$.}
\end{center}
\end{figure}

	Following~\cite{oetal98}, the averaged
gravitational wave amplitude is then given by
\begin{equation}
\label{h_ts}
h(t) = \sqrt{\frac{1}{10} \langle\langle h_+^2
	+h_\times^2\rangle\rangle} =
	\frac{256}{135}\sqrt{\frac{\pi}{30}} 
	\frac{G M}{c^5} \frac{\alpha\Omega^3 R^3}{D}\tilde{J}
	\ ,
\end{equation}
and its time evolution for a ``fiducial'' neutron star
placed at 20 Mpc is shown in Fig.~\ref{fig7}. The left
panel, in particular, shows evolutionary tracks for a
number of different initial values of the magnetic field
and a saturation amplitude $\alpha_{\rm sat}=1$. The
tracks terminate when $g(t)=1$, which is marked with an
open symbol (different symbols refer to different initial
magnetic fields). The right panel, on the other hand,
refers to an initial magnetic field $\langle B_0
\rangle=10^{10}$ G and shows the effect of different
saturation amplitudes on the reaching of a magnetic
energy production rate sufficient to suppress the
instability (which occurs at the point marked with filled
squares). Taking a saturation amplitude $\alpha_{\rm
sat}=0.1$, one sees that the emission of gravitational
waves would start being suppressed after about 20 days.

	A more useful quantity for discussing the
detectability of the gravitational waves emitted is the
characteristic amplitude $h_c$, defined as
\begin{equation}
h_c(f) \equiv h(t) \sqrt{f^2
	\bigg\vert\frac{dt}{df}\bigg\vert} \ ,
\end{equation}
where $f$ is the gravitational wave frequency and $f =
\omega/(2\pi) = 2\Omega/(3\pi)$ for a $\ell=2$ mode. 

\begin{center}
\begin{figure}[htb]
\leavevmode
\psfig{file=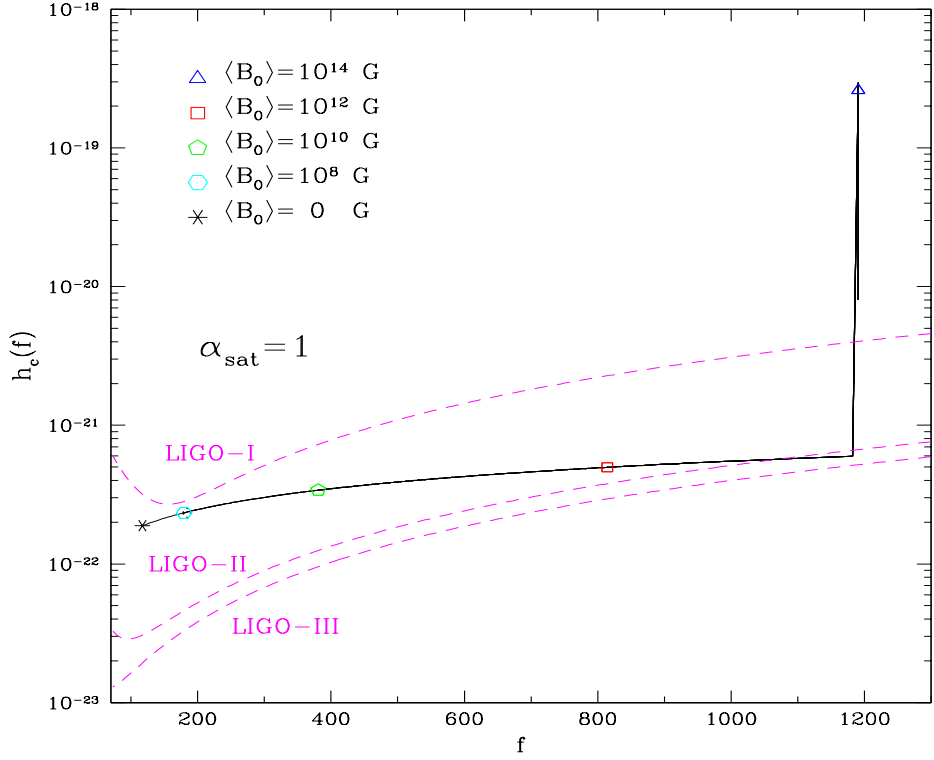,angle=0,width=13.0truecm,height=11.0truecm}
\caption[]{\label{fig8} Frequency evolution for the
characteristic amplitude for a magnetic neutron star
located at 20 Mpc and with a saturation amplitude
$\alpha_{\rm sat}=1$. As in Fig.~\ref{fig7}, different
symbols mark the frequency at which $g(t)=1$ for
different values of the initial magnetic field. The three
dashed lines show the expected $h_{\rm rms}$ for the
three stages of development of the LIGO interferometers.
Note that the spike in $h_c$ is the result of a vanishing
$df/dt$ at the onset of the instability and does not
contribute significantly to the SNR.}
\end{figure}
\end{center}

	In Fig.~\ref{fig8} we show the frequency
evolution of the characteristic amplitude for a magnetic
neutron star located at 20 Mpc and with a saturation
amplitude $\alpha_{\rm sat}=1$\footnote{Because
$h(t)\propto \alpha_{\rm sat}$, but $|dt/df|\propto
\alpha^{-2}_{\rm sat}$, the characteristic amplitude is
basically insensitive to $\alpha_{\rm sat}$}. As in
Fig.~\ref{fig7}, different symbols mark the frequency at
which $g(t)=1$ for different values of the initial
magnetic field. The frequency evolutionary track should
be compared to the expected root-mean-square strain noise
$h_{\rm rms}$ curves, indicated with dashed lines
Fig.~\ref{fig8} and referring to the LIGO ``first
interferometers''~(LIGO~I), to the ``enhanced
interferometers''~(LIGO~II), and to the ``advanced
interferometers''~(LIGO~III). Analytic expressions for
the expected $h_{\rm rms}$ have been presented
in~\cite{oetal98}.

	The characteristic amplitude also allows a direct
calculation of the signal-to-noise ratio (SNR) for a
single source at a distance $D$. Using matched filtering
techniques, the SNR can be written as~\cite{oetal98}
\begin{equation}
\label{s_o_n}
\left({S\over N}\right)^2
	=2\int^{f_{\rm max}}_{f_{\rm min}}
	{df\over f}\left({h_c\over h_{\rm rms}}\right)^2 \ ,
\end{equation}
where, for a strain noise with power spectrum $S_h(f)$,
\hbox{$h_{\rm rms} \equiv \sqrt{f S_h(f)}$.} The integral
in (\ref{s_o_n}) is computed between the minimum
frequency $f_{\rm min}= 2 \Omega_0/(3 \pi)$ and the
maximum frequency $f_{\rm max}= 2 \Omega_{\rm crit}/(3
\pi)$ at which the gravitational radiation is emitted,
but is largely dominated by the amplitude-saturated part
of the evolution. As a consequence, it is strongly
influenced by the angular velocity at which the mode is
suppressed $\Omega_{\rm crit}$ and, as discussed in the
previous Section, on the initial strength of the magnetic
field.

	In Table II we report the results of the
calculation of signal-to-noise ratios obtained through
the evolutionary models discussed in the previous
Sections. The SNR are given for $\alpha_{\rm sat}=1$,
($\alpha_{\rm sat}=10^{-1}$) and different values of the
initial magnetic field. For a zero initial magnetic field
and $\alpha_{\rm sat}=1$, we recover the results obtained
by~\cite{oetal98}

\vbox{
\begin{table}[htb]
\begin{tabular}{|c|ccccc|}
         &        &         &       &     & 		\\
 $S/N: \alpha_{\rm sat}=1$ & $\langle B_0 \rangle = 0      \ $G \ \ 
	 & $\langle B_0 \rangle = 10^8   \ $G \ \ 
	 & $\langle B_0 \rangle = 10^{10}\ $G \ \ 
	 & $\langle B_0 \rangle = 10^{12}\ $G \ \ 
	 & $\langle B_0 \rangle = 10^{14}\ $G \ \ 			\\ 
 $(\alpha_{\rm sat}=10^{-1})$&     &           &          &            &\\
         &            &           &          &            &	\\
\tableline
         &            &           &           &           &	\\
LIGO-I   & 1.2 (0.7)  & 0.8 (---) & 0.4 (0.4) & 0.1 (0.1) & $< 0.01$\\
LIGO-II  & 7.6 (4.1)  & 4.8 (4.1) & 2.2 (2.1) & 0.8 (0.6) & $< 0.01$\\
LIGO-III & 10.6 (5.4) & 6.4 (5.4) & 2.9 (2.7) & 1.1 (0.8) & $< 0.01$\\ 
         &            &           &           &           &	\\
\end{tabular}
\label{table2}
\bigskip
\caption[]{Signal-to-Noise ratios for different values of
the saturation amplitude $\alpha_{\rm sat}$ and of the
initial magnetic field $\langle B_0 \rangle$. The values
referring to a zero initial magnetic field coincide with
those calculated by~\cite{oetal98}. As for Table I, the
values not reported refer to times longer than one year.}
\end{table}}

	Note that the presence of a relatively weak
initial magnetic field (i.e. $\langle B_0 \rangle = 10^8$
G) does not significantly change the SNR as compared to
an neutron star with zero magnetic field. However, this
is not the case when the initial magnetic field is taken
to be stronger and, in particular, for an initial
$\langle B_0 \rangle = 10^{12}$ G, the SNR is such that
even LIGO-III would only be able to detect it if
$\alpha_{\rm sat}$ is ${\cal O}(1)$. Overall our results
indicate therefore that the development of the $r$-mode
instability in a magnetic neutron star could
significantly reduce the expectations of detection of
gravitational waves from $r$-mode oscillations.

\section{Conclusions}
\label{conclusions}

	We have computed numerically the growth of the
magnetic field generated by the secular kinematic effects
emerging during the evolution of the $r$-mode
instability. We have treated the magnetic field growth as
a kinematic dynamo problem, thus neglecting any
back-reaction of the magnetic field on the forces driving
the velocity field. The toroidal magnetic field computed
in this way exhibits an exponential growth which is a
direct consequence of the exponential development of the
$r$-mode instability. During this stage of the growth,
the toroidal field can become extremely large and, in the
case of a saturation amplitude of order unity, it can
grow to be four orders of magnitude larger than the seed
poloidal field. At mode saturation and thereafter, the
field growth rate is smaller but nonzero, thus still
contributing significantly to the production of very
intense magnetic fields.

	Although we have neglected the back-reaction of
the magnetic field on the kinematics of the $r$-mode
oscillations, we can estimate the strength of the
magnetic fields that would either prevent the onset of
the \hbox{$r$-mode} instability or damp it when this is
allowed to grow. Our estimates indicate that the critical
initial magnetic field which would suppress even the
first $r$-mode oscillation is $\langle B_0\rangle_{\rm
crit}\sim 10^{16}$ G and this is basically independent of
whether the core is superconducting or not. Such a
magnetic field is two orders of magnitude larger than
those usually associated with young neutron stars and it
is therefore likely that most newly born neutron stars
will have an initial magnetic field which is sufficiently
weak for the instability to be triggered. However,
depending on the strength of the initial magnetic field
and on the mode amplitude at saturation, an
\hbox{$r$-mode} instability which is initially free to
develop could be subsequently damped by the generation of
magnetic fields on timescales much shorter than the ones
set by viscous effects. As an example, with an initial
magnetic field strength $\langle B_0 \rangle = 10^{10}$ G
and a normal stellar core, we find that the instability
would be suppressed in less than four hours for a
saturation amplitude $\alpha_{\rm sat}=1$ and in less
than ten days for $\alpha_{\rm sat}=0.1$.

	An important point to stress is that even in the
case the $r$-mode instability is rapidly suppressed by
the coupling of the mass-currents with magnetic fields,
the star will have had sufficient time to loose a good
fraction of its rotational energy to gravitational
waves. This is because the loss-rate of angular momentum
to gravitational waves is a steep function of the star's
angular velocity and is therefore most efficient only in
the very initial stages of the instability. As a result,
for realistic values of the initial magnetic field, most
of the astrophysical considerations about the importance
of the $r$-mode instability in spinning down newly-born
neutron stars remain basically valid.

	The validity of our scenario could be verified in
a number of ways. As discussed in the previous Section,
the larger the initial magnetic field, the sooner the
condition for mode damping will be reached. We could
therefore expect a positive correlation of the stellar
spin rate at about one year and the initial poloidal
magnetic field of the neutron star. For the same reason,
we could expect an anti-correlation between the stellar
spin rate and the toroidal magnetic field at about one
year. Our results also have a direct impact on the
gravitational wave detectability of $r$ modes from
unstable neutron stars. Since the SNR for the detection
of gravitational waves from \hbox{$r$-mode} oscillations
depends sensitively on the star's spin frequency at mode
suppression, and the latter is determined by the mode
amplitude at saturation and the initial magnetic field,
we have found that the SNR could be significantly above
unity only if the initial magnetic field is below
$10^{12}$ G. Moreover, because of the quantitative
difference in the mode damping in the case of a
superconducting stellar core, the determination of the
critical frequency could be used to infer a signature of
the onset of proton superconductivity in the core.

	Finally, we note that some fundamental issues
remain unresolved. A most outstanding one is whether the
magnetic field will distort the $r$-mode velocity field
in such a fine-tuned way so as to cancel the fluid drift
before it suppresses the oscillation. As discussed
in~\cite{rls00}, considerations of the energy densities
in play suggest that this is rather unlikely, but a more
detailed analysis of this would certainly provide
important information about the evolution of the
\hbox{$r$-mode} instability in magnetic neutron
stars.

\acknowledgments We are grateful to Nils Andersson,
Kostas Kokkotas, John Miller, Nikolaos Stergioulas and
Shin'ichirou Yoshida for the numerous discussions and for
carefully reading the manuscript. LR also thanks Scott
Hughes for the useful conversations on gravitational wave
detection. LR acknowlodges support from the Italian MURST
and by the EU Programme "Improving the Human Research
Potential and the Socio-Economic Knowledge Base"
(Research Training Network Contract
HPRN-CT-2000-00137). FKL, DM and SLS acknowledge support
from the NSF grants AST~96-18524 and PHY~99-02833 and
NASA grants NAG~5-8424 and NAG~5-7152 at Illinois. FKL is
also grateful for the hospitality extended to him by John
Miller, SISSA, and ICTP, where this work was completed.

\end{document}